\documentclass[useAMS,usenatbib]{mn2e}
\usepackage{epsfig}

\title{The ratio of pattern speeds in double-barred galaxies}
\author[J. Font, J.E. Beckman, J. Zaragoza-Cardiel, K. Fathi, B. Epinat \& P. Amram ]{Joan Font$^{1}$\thanks{E-mail:
jfont@iac.es}John E. Beckman$^{1,2,3}$, Javier Zaragoza-Cardiel$^{1,2}$, Kambiz Fathi$^{4,5}$, \and
Benoit Epinat$^{6,7}$, Philippe Amram$^{6,7}$ \\
$^1$Instituto de Astrof\'\i sica de Canarias, c/ V\'\i a L\'actea s/n, E38205, La Laguna, Tenerife, Spain\\
$^2$Departamento de Astrof\'\i sica. Universidad de La Laguna, Tenerife, Spain\\
$^3$Consejo Superior de Investigaciones Cient'ficas, Spain\\
$^4$Department of Astronomy, Stockholm University, AlbaNova Center, 106 91 Stockholm, Sweden\\
$^5$Oskar Klein Centre for Cosmoparticle Physics, Stockholm University, 106 91 Stockholm, Sweden\\
$^6$Laboratoire dÕAstrophysique de Marseille, Universit\'e dÕAix-Marseille. Marseille. France\\
$^7$CNRS, UMR7326, 38, rue F. Juliot-Curie, 13388 Marseille Cedex 13, France}
\begin{document}

\date{Draft \today}
\pagerange{\pageref{firstpage}--\pageref{lastpage}} \pubyear{2014}

\maketitle

\label{firstpage}

\begin{abstract}
We have obtained two-dimensional velocity fields in the ionized gas of a set of 8 double-barred
galaxies, at high spatial and spectral resolution, using their H$\alpha$ emission fields
measured with a scanning Fabry-Perot spectrometer. Using the technique by which phase
reversals in the non-circular motion indicate a radius of corotation, taking advantage of the
high angular and velocity resolution we have obtained the corotation radii and the pattern
speeds of both the major bar and the small central bar in each of the galaxies; there are few
such measurements in the literature. Our results show that the inner bar rotates more
rapidly than the outer bar by a factor between 3.3 and 3.6.
\end{abstract}


\section{Introduction}

 The idea that nested bars could provide a mechanism for feeding the central massive black holes of galaxies dates from an article by \citet{b44}. It was understood that a major bar in a galaxy allows, via angular momentum exchange, gas to flow towards the center, but when this in gas reaches a galactocentric radius of some the half the bar width the gravitational asymmetry reaches virtually zero, so the mechanism is no longer effective. \citet{b44} postulated that a small, nuclear bar within the main bar, offering an asymmetric field allows inflow to continue towards the nucleus.  As observations increased in sensitivity, a significant fraction of barred galaxies was observed to harbour secondary nuclear bars, (see \citet{b14,b15}) although there may not be a preponderance of double barred galaxies among active galaxies \citep{b16}. 
 
 It is important to know whether nuclear bars are dynamically decoupled from the outer bar. Early simulation arguments \citep{b44, b37} suggested that they should be decoupled and rotating faster than the outer bars. This received support from orbital models \citep{b39, b31, b40, b32} and N-body simulations \citep{b40,b43}. Indirect evidence that the bars are decoupled is that that their relative position angles are randomly distributed \citep{b4,b15}. Models of individual galaxies where the gas morphology is derived from a potential model with two rotating bars also indicate that the two should rotate with different pattern speeds; e.g. NGC 4314 \citep{b1} and NGC1068 \citep{b13}.  For the latter the authors used supporting kinematic measurements from a Sauron data cube.
 
 There are few measurements in the literature of pattern speeds for nuclear bars in double-barred galaxies. \citet{b7} applied the Tremaine-Weinberg (TW) method \citep{b47} to the stellar component of NGC 2950, showing that the two have different pattern speeds, although it was difficult to obtain a reliable value for the inner bar. \citet{b18} used the TW method on the H$\alpha$ emission line velocity fields of 10 galaxies, finding evidence in three of them for a more rapidly rotating nuclear bar. As shown by \citet{b33}, the TW method may not give a true value for the inner pattern speed, but it can show that there is an inner component in rotation at a higher angular velocity than the outer component.
 
We have developed a new general method for finding corotation radii, and hence pattern speeds of resonant systems in disc galaxies. This relies on measuring, at high enough angular and velocity resolution, the radii at which the non-circular velocity of the gas changes phase \citep{b23,b6}. We use 2D velocity maps in H$\alpha$ emission obtained with a Fabry-Perot spectrometer. This method requires sufficient star forming activity to produce widespread H$\alpha$, which allows us to measure late type disc galaxies. It was introduced in \citet{b19}, applied to a sample of eight galaxies; in \citet{b20} we applied an improved version to 104 galaxies. The angular resolution available, in the range 0.8 - 4.4 arcsec, lets us find the corotations of nuclear bars as well as major bars and the spiral arm system. Here we have applied it to eight double barred galaxies.
 
In the next section we describe briefly the method, in section three we give the sources of our observations, and present our results, including analysis of the resonances in the individual galaxies; in section four we discuss our results.

\section{The Streaming Phase-Reversal Method}

	The principles of the method used were described in \citet{b20}. Here we give a summary, to explain the guidelines of the technique. Using a Fabry-Perot data cube in H$\alpha$ emission we derive a spectrum for each position on a galaxy, and transform it into 2D maps of integrated flux, l.o.s. velocity and velocity dispersion. Deriving the rotation curve from the velocity field and subtracting it off, in two dimensions, from the original map we get the map of the residual, non-circular (ÒstreamingÓ) velocities. We effect the separation of circular and non-circular velocity over the full 2D field, using an iterative process to yield a non-axisymmetric field, which becomes the prime material for our method.
	
\begin{figure*}
\begin{center}
  \includegraphics[width=.9\textwidth]{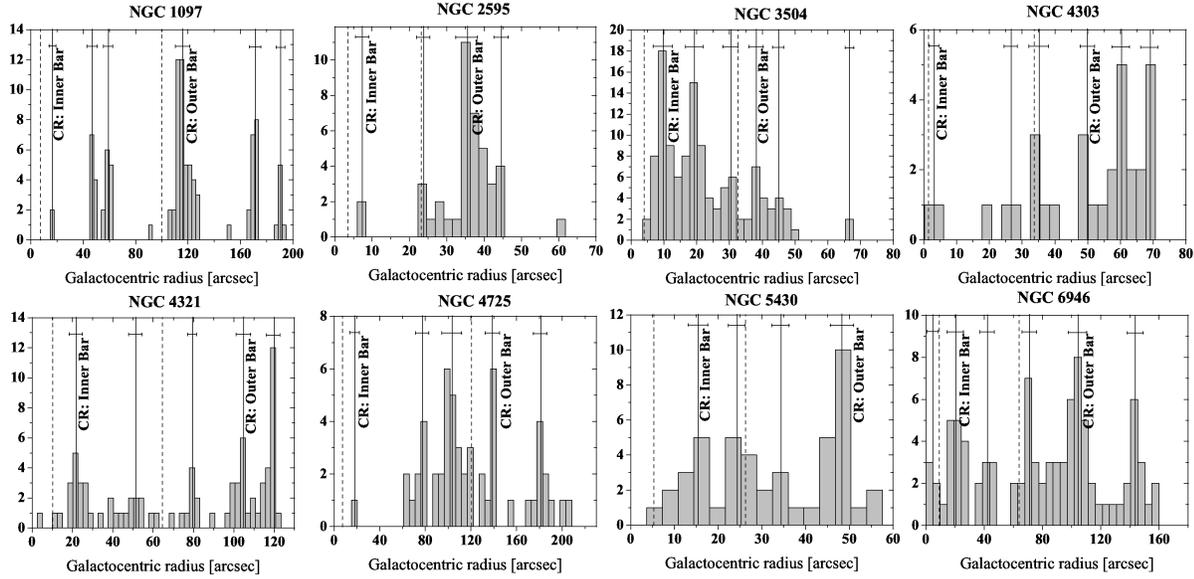}
   \caption{Histograms of the phase reversals as a function of the radial distance, for each galaxy. Ordinate axis shows the number of phase reversals obtained using the residual velocity maps. Dashed vertical lines indicate lengths of nested bars. Solid vertical lines mark the central positions of those peaks identified as resonances; the horizontal segments show the uncertainty in the radius of the resonance.} 
  \label{fig:histos}
\end{center}
\end{figure*}

	This residual velocity map is used to derive the phase reversal histogram (see figure \ref{fig:histos}). We locate those positions in the residual velocity field in which the streaming velocity changes sign when traced along the radial direction (a phase-reversal). To minimize noise and projection problems, a valid phase-reversal must satisfy two criteria, based on the angular and the velocity resolutions  \citep{b20}. Plotting the radial distribution for the phase-reversals as a function of galactocentric radius we assign each of the maxima in this histogram to a resonance. From the rotation curve, we can derive the frequency curves $\Omega, \Omega\pm\kappa/2, \Omega\pm\kappa/4$, where $\Omega$ is the angular speed and $\kappa$ the epicyclic frequency. These curves are used to classify the resonance as corotation, inner and outer Lindblad resonance (ILR, OLR), plus the ultraharmonic resonances (UHR), to study the couplings between the resonances, and to determine the pattern speeds. In  \citet{b20} we found that most resonances for most of the galaxies are associated with corotations, with a small fraction of objects showing peaks which can be identified as  ILRÕs or OLRÕs. 
	
\section{Results}
\subsection{The observational data}

The data set for three of the galaxies in our sample was the GHASP Fabry-Perot database \citep{b10} taken on the 1.93m telescope at the Observatoire de Haute Provence. We selected those galaxies where we could see a double bar: a large outer bar and a small central bar. The galaxies chosen were:  NGC 2595, NGC 5430 and NGC6946. The moment maps of these objects are available on the data base \citep{b45}. The observational parameters for these objects: as pixel scale, angular and spectral resolution, and geometrical parameters (position angle, inclination and distance) can be found in \citep{b10}.

We take from the same data base, the first moment maps of NGC4321, NGC1097 and NGC4725, the first from the VIRGO survey \citep{b5} using the FaNTOmM Fabry-Perot at the1.6m Mont MŽgantic telescope, these authors also give the observational and geometrical paramters for this galaxy; the two remaining galaxies belong to the SINGS data base \citep{b24}; \citep{b8,b51} took the Fabry-Perot observations, respectively, at Mont MŽgantic,  The geometrical parameters for NGC 1097 can be found in  \citet{b38}, while the observational parameters for this galaxy together with all parameters of NGC 4725 are given in \citet{b8}.

The data cubes for M61 (NGC 4303) and NGC 3504, obtained at the 4.2m William Herschel telescope, La Palma, with the GH$\alpha$FaS Fabry-Perot instrument, show the highest angular resolution or "seeing" value (1.2 and 0.8 arcsec, respectively) and spectral sampling resolution (8.25 km$\cdot$s$^{-1}$) , and S:N ratio.

\subsection{Corotations and associated resonance radii}

	In Figure \ref{fig:histos} we show, for each galaxy, a histogram giving the radial dependence of the number of detected phase reversals. The vertical solid lines mark the radial positions of the maxima in the distributions, positively identified as resonances; the horizontal bars at the top show the uncertainty for each resonance. In the present study, we only consider the corotations associated with the two bars. To establish which of the resonances corresponds to the corotation of a given bar, we first apply the bar corotation (BC) criterion defined in \citet{b20}: the strongest peak in the phase-reversal distribution located near the end of the bar is assumed to lie at its corotation. The position of this peak gives the bar corotation radius. When we find two or more peaks which satisfy the BC criterion we use the predictions by  \citet{b53} who give, for the main bar, values for $\mathcal{R}$, the ratio of the corotation radius to the bar length as a function of morphological type, to decide which peak corresponds to the bar corotation radius. In each histogram of Figure \ref{fig:histos}, the corotations of the nuclear bar and of the outer bar are labeled as ÒCR:Inner BarÓ and ÒCR:Outer BarÓ, respectively. The dashed vertical lines show the lengths of the bars, given in Columns (4) and (8) of  Table 1. We now present a brief description of the resonances for each galaxy.

\begin{table*}
 \begin{minipage}{180mm}
  \caption{Dynamical parameters of nested bars.}
  \begin{tabular}{@{}ccccccccccc@{}}
  \hline
   {Object}    &  Type & {r$^{in}_{bar}$} & {r$^{in}_{CR}$}  &  {$\mathcal{R}^{in}$}  &  {$\Omega^{in}$}  &  {r$^{out}_{bar}$}  &  {r$^{out}_{CR}$}  &  {$\mathcal{R}^{out}$} &  {$\Omega^{out}$}   &  {$\Gamma$}  \\
    {NGC}   &   & {(arcsec)} &  (arcsec) &    &  {(km$\cdot$s$^{-1}\cdot$kpc$^{-1}$)} & {(arcsec)}  & (arcsec) &  &  {(km$\cdot$s$^{-1}\cdot$kpc$^{-1}$)}  & \\
       {(1)} & {(2)}  & {(3)}  & {(4)} & {(5)} & {(6)} & {(7)} & {(8)}  & {(9)} &  {(10)}  & (11)\\

 \hline
{1097} & SB(s)b  &{7.2}                  & {16.5$\pm$2.7} & {2.3$\pm$0.4} & {136.8$^{+12.8}_{-9.6}$} & {98.2$\pm$14.4} & {114.5$\pm$7.1} & {1.1$\pm$0.2} & {38.3$^{+2.3}_{-2.1}$} & {3.6$^{+0.4}_{-0.3}$}\\
{2595} & SAB(rs)c  &{4.2$\pm$0.2} & {8.9$\pm$1.2} & {2.1$\pm$0.3} & {123.2$^{+15.9}_{-12.5}$} & {23.8$\pm$1.0} & {36.5$\pm$2.4} & {1.5$\pm$0.1} & {35.1$^{+2.4}_{-2.1}$} & {3.5$^{+0.5}_{-0.4}$}\\
{3504} & SAB(s)ab  &{4.1$\pm$0.4} & {9.9$\pm$2.7} & {2.4$\pm$0.7} & {230.5$^{+104.7}_{-59.9}$} & {33$\pm$3} & {37.6$\pm$2.4} & {1.2$\pm$0.1} & {67.1$^{+3.5}_{-3.3}$} & {3.4$^{+1.5}_{-0.9}$}\\
{4303} & SAB(rs)bc  &{2.1$\pm$0.1} & {4.6$\pm$1.0} & {2.2$\pm$0.4} & {135.0$^{+10.0}_{-8.4}$} & {33.0$\pm$2.6} & {49.9$\pm$1.1} & {1.5$\pm$0.1} & {39.0$^{+0.7}_{-0.6}$} & {3.5$^{+0.5}_{-0.4}$}\\
{4321} & SAB(s)bc  &{9.7$\pm$0.5} & {20.9$\pm$3.3} & {2.2$\pm$0.4} & {81.1$^{+11.3}_{-8.6}$} & {72.0$\pm$12.0} & {102.5$\pm$3.3} & {1.4$\pm$0.1} & {22.8$^{+0.7}_{-0.6}$} &  {3.5$^{+0.5}_{-0.4}$}\\
{4725} & SAB(r)ab  &{8.2$\pm$0.8} & {17.8$\pm$3.8} & {2.2$\pm$0.5} & {92.3$^{+7.9}_{-6.6}$} & {123.3$\pm$5.2} & {139.7$\pm$3.8} & {1.2$\pm$0.1} & {26.7$^{+0.7}_{-0.6}$} & {3.5$^{+0.3}_{-0.3}$}\\
{5430} & SB(s)b  &{5.6$\pm$0.6} & {15.2$\pm$3.0} & {2.7$\pm$0.8} & {79.5$^{+17.1}_{-12.4}$} & {26.1$\pm$1.0} & {47.5$\pm$2.6} & {1.8$\pm$0.1}& {23.8$^{+1.7}_{-1.5}$} & {3.3$^{+0.8}_{-0.6}$}\\
{6946} & SAB(rs)cd  &{9.2$\pm$1.2} & {20.2$\pm$4.8} & {2.2$\pm$0.6} & {174.9$^{+36.7}_{-24.8}$} & {65$\pm$5} & {102.9$\pm$6.6} & {1.6$\pm$0.2} & {51.3$^{+2.8}_{-2.5}$} &{3.4$^{+0.7}_{-0.5}$}\\
\hline
\end{tabular}
\\Column (1): names of objects, Column (2) morphologycal type taken from RC3 \citep{b60}. Column (3) nuclear bar radius (see text for details and references). Columns (4), (5), respectively:  corotation radius of nuclear bar from our phase reversals method, and $\mathcal{R}$: ratio of the corotation radius and the bar length. Column (6): pattern speed value associated with bar corotation.  Columns (7)-(10):  same parameters as columns (3)-(6) but for the large bar. Column (11): values of {$\Gamma$} the ratio between the inner and outer pattern speeds.
\end{minipage}
\end{table*}

\subsection{Resonance structures in the individual galaxies}

\textbf{NGC1097}. This nearby SBb galaxy (D = 14.5 Mpc) has a nested bar system, first reported by \citet{b3}, and later measured by \citet{b49}. The two bar radii are given in  Table 1 (columns 3, 7). The inner bar length is from \citet{b28}, and from \citet{b9} who argue that the bar length should be less than 8 arcsec. The outer bar length is the mean of the values given in \citet{b49}, \citet{b28}, \citet{b35} and \citet{b38}; the associated uncertainty is the standard deviation of these values. The histogram (in Fig.  \ref{fig:histos}) shows six peaks. Peak four, the strongest is assigned to the corotation radius of the main bar using the BC criterion, and we take peak one as corotation of the inner bar. These radii are in  Table 1, columns 4 and 8. Our values for the corotation radius of the outer bar (114.5$\pm$7.1 arcsec) and its angular rate (38.3$\pm$2.2 km$\cdot$s$^{-1}\cdot$kpc$^{-1}$), agree, within the uncertainties, with previous results obtained using different methods (see   Table 1, columns 8 and 10). \citet{b48}, find an angular rate and a corotation radius of and 35 km$\cdot$s$^{-1}\cdot$kpc$^{-1}$ and 122.3 arcsec, respectively. Both values are reproduced by \citet{b38}, using a dynamical model for the gravitational potential to determine the main bar pattern speed as 36$\pm$2 km$\cdot$s$^{-1}\cdot$kpc$^{-1}$, placing corotation at a deprojected distance of 122.3$\pm$7.1 arcsec from the galaxy center.  They also apply the Tremaine-Weinberg method to an ionized gas velocity field to calculate the pattern speed, finding a value of 30$\pm$8 km$\cdot$s$^{-1}\cdot$kpc$^{-1}$.

\textbf{NGC2595}. This galaxy has a clear double bar. The assumed distance is 58.1 Mpc. Our procedure yields four peaks (Figure \ref{fig:histos}). Applying the BC criterion peaks 1 and 3 lie at corotation for the inner and outer bar corotation, respectively. The radii of the two bars are estimated by finding the radial position of the maxima in ellipticity at constant position angle for ellipse fitting of the isophotes on a 3.6 $\mu$m image (Spitzer archive), which are deprojected using the PA of the bar and PA and inclination of the disc. 

\textbf{NGC3504}. In this barred, ringed galaxy \citep{b4} we find six peaks (Figure \ref{fig:histos}). We attribute peak 1 to the corotation of the inner bar and peak 4 to that of the outer bar, using the BC criterion. The outer bar length is from \citet{b25}; the semimajor axis of the nuclear bar is determined from the ellipticity and position angle curves of the isophotes as functions of radius. Our pattern speed for the main bar ($\Omega_{P}$ = 67.6 km$\cdot$s$^{-1}\cdot$kpc$^{-1}$ with D = 19.8 Mpc) is quite close to $\Omega_{P}$ = 77 km$\cdot$s$^{-1}\cdot$kpc$^{-1}$ found by \citet{b25}, but inconsistent with the upper limit $\Omega_{P}\leq$ 41 km$\cdot$s$^{-1}\cdot$kpc$^{-1}$, determined by \citet{b30}; this is satisfied only by our outermost resonance at 66.6$\pm$1.6 arcsec (see Figure  \ref{fig:histos}) with an angular speed of 37.9$\pm$0.9 km$\cdot$s$^{-1}\cdot$kpc$^{-1}$.

\textbf{NGC 4303 (M61)}. We find six peaks in the histogram (Figure  \ref{fig:histos}). Using the BC criterion we find corotation for the inner bar and the main bar as peaks 1 and 4, respectively. The inner bar length is the mean of those of \citet{b14} and \citet{b36}, while the outer bar length is the value of \citet{b14}. \citet{b27} apply a cloud-orbit model finding a corotation radius of 38.5 arcsec for the main bar, almost compatible with our value (Table 1). Our corotation radius of the nuclear bar (4.6 arcsec) is consistent with the length of the gas bar in \citet{b27}, and with the analysis of \citet{b42}, who argued that it should lie between the inner ILR (at 2 arcsec) and the outer ILR (at 10-14 arcsec) of the outer bar. \citet{b52} establish a lower limit for the primary bar corotation radius $r_{CR} \geq$ 36 arcsec (consistent with our result), and estimate the angular speed of the outer bar assuming a distance of 16.1 Mpc, $\Omega_{P}$ = 24$\pm$29 km$\cdot$s$^{-1}\cdot$kpc$^{-1}$, compatible with our value given the large uncertainty.

\textbf{NGC 4321 (M100)}. This large nearby spiral (D = 15.9 Mpc) is intensively studied. Its inner bar is clearly visible in the near IR (see \citet{b26}), but its major bar is weak, and rather oval in shape. We find five peaks in our histogram (Figure \ref{fig:histos}). The BC criterion yields peak 1 as the corotation of the inner bar and peak 4 that of the outer bar. The sizes of the bars are from \citet{b14}, noting also the value by \citet{b35}. The corotation radius for the main bar determined here and its pattern speed (see   Table 1) are in agreement with most previous studies which used a wide variety of methods. We have discussed these, for the outer bar only, in \citet{b19}, to which the reader is referred. \citet{b22} used the TW method to determine the pattern speed of the inner bar, finding 48 km$\cdot$s$^{-1}\cdot$kpc$^{-1}$ which not unexpectedly using TW on gas for an inner bar, does not coincide with our value.

\textbf{NGC 4725}. This early type ringed galaxy shows a diffuse primary bar with a nested nuclear bar. The phase reversals show five peaks (Figure \ref{fig:histos}). With the BC criterion we assign peaks 1 and 4 respectively to the corotations of the inner and outer bars. The bar lengths are from \citet{b14} and \citet{b35}. \citet{b3} found a pattern speed for this galaxy between 13.1-29.5 km$\cdot$s$^{-1}\cdot$kpc$^{-1}$, placing corotation at 186 arcsec. These are not compatible with our values for the outer bar (Table 1), but agree with the outermost resonance in our phase histogram, which we attribute to the spiral structure, and is located at 181.3$\pm$4.9 arcsec, with an angular speed of 21.1$\pm$0.6 km$\cdot$s$^{-1}\cdot$kpc$^{-1}$ (taken a distance of 17.1 Mpc).

\textbf{NGC 5430}. This galkaxy, located at 49.0 Mpc, has a strong outer bar, with asymmetric outer disc, suggesting a recent merger. It shows four histogram peaks. Using the BC criterion we assign peak 1 to the nuclear bar corotation, and peak 2 to that of the outer bar. The length of the nuclear bar is determined as the radius of the local maximum of the ellipticity curve, while the radius of the main bar is calculated by deprojecting the value given in \citet{b21}.

\textbf{NGC6946}. \citet{b11} first reported the bar-within-a-bar morphology, analysing near-infrared images of this late-type galaxy. The histogram, within a region of radius 125 arcsec, shows six peaks. Peaks two and five are assigned as corotations of the two bars, using the BC criterion.The outer bar size ( Table 1)  is an average of the measurements of \citet{b41} and of \citet{b11}; for the nuclear bar we calculated the mean of the values of \citet{b11} and \citet{b46}. The TW method was applied to CO data by \citet{b50} and to H$\alpha$ data by \citet{b17}. The first authors found a pattern speed of 39$\pm$13 km$\cdot$s$^{-1}\cdot$kpc$^{-1}$, which is well reproduced if corotation occurs at the outermost peak (see Figure \ref{fig:histos}), at 143.6±3.3 arcsec for which the angular velocity is 39$\pm$13 km$\cdot$s$^{-1}\cdot$kpc$^{-1}$ (assuming a distance of 5.9 Mpc). \citet{b17} determine two pattern speeds: a primary one, of 22$^{+4}_{-1}$ km$\cdot$s$^{-1}\cdot$kpc$^{-1}$, at a radius of 304 arcsec (beyond the limits of our data) and a secondary one of 47$^{+3}_{-2}$ km$\cdot$s$^{-1}\cdot$kpc$^{-1}$, which is in agreement, within uncertainties, with the angular rate for the outer bar found here. Values for the corotation radius and pattern speed by authors using other techniques can be found in \citet{b17}. 
	
\section{discussion}

We have determined the pattern speeds of the major bar and the nuclear bar in eight galaxies, from early to late types, using emission in H$\alpha$ from the ionized gas in their interstellar media. The technique is straightforward, but can be performed only with 2d coverage of the velocity field across the whole galaxy disc, with good angular and velocity resolution. This is possible using Fabry-Perot observations as described here, but would be possible using maps of sufficient resolution in HI or CO emission lines. We find in all cases that the nuclear bar is rotating with angular velocity considerably greater than the outer bar, showing that the bars are dynamically decoupled. The range of values for the pattern speed ratio is quite narrow, ranging from 3.3 to 3.6, with a mean value of 3.5$^{+0.7}_{-0.5}$. This result is in good agreement with gas models by \citet{b12} who predicted a value for this ratio of 3.4 once the inner bar has stabilized its rotation. They showed that before it has achieved this stability the ratio should oscillate between 3.2 and 3.8, which fits surprisingly well the range of values we find here. A series of models in \citet{b32} predict values for the ratio of bar angular rates between 1.6 and 2.7. We note also that the lowest value for the pattern speed of an inner bar, 79.5 km$\cdot$s$^{-1}\cdot$kpc$^{-1}$, is quite close to the lower limit suggested in \citet{b34} for orbitally stable stellar bars.
The ratio between the corotation radius of the outer bar and its length is greater than 1 (Table 1, column 9), agreeing with simulations by \citet{b2}. The values for this ratio agree particularly well with the model predictions of  \citet{b53}, these authors find that this ratio depends on the morphological type, increasing from early type ($\mathcal{R}$=1.15$\pm$0.25) to intermerdiate type ($\mathcal{R}$=1.44$\pm$0.29) and late type galxies ($\mathcal{R}$=1.82$\pm$0.63). The equivalent ratio for the inner bar (Table 1, column 5) is found to be between 2.1 and 2.7 with a mean of 2.3, agreeing with predictions by \citet{b34}.

\section*{Acknowledgments}

GH$\alpha$FaS is a visitor instrument on the William Herschel Telescope, Isaac Newton Group, IAC Observatorio del Roque de los Muchachos, La Palma. We thank Miguel Querejeta for data analysis contributions, Phil James for observations of NGC 4303. J.E.B. thanks APCTP for hospitality at the 7th Korean Astrophysics Workshop; KF thanks the Swedish Research Council and the Swedish Royal Academy of Sciences Crafoord Foundation. Support comes from project AYA2007-67625-CO2-01 of the Spanish Ministry of Science and Innovation, project P3/86 of the Instituto de Astrof'sica de Canarias, and the DAGAL network, Marie Curie actions,EU 7th framework programme FP7/2007-2013/REA grant agreement PITN-GA-2011-289313.

\bsp

\label{lastpage}

\end{document}